\documentclass[twocolumn,prd,superscriptaddress,showpacs,amsmath,amssymb]{revtex4-1}

\usepackage{graphicx}
\usepackage{dcolumn}
\usepackage{bm}
\usepackage{multirow}
\usepackage{enumitem}

\begin{document}

\title{Search for a Non-Relativistic Component in \\ the Spectrum of Cosmic Rays at Earth}

\def\UC{Enrico Fermi Institute, Department of Physics, and Kavli Institute for Cosmological Physics, University of Chicago, Chicago, Illinois 60637, USA}

\author{J.I.\!\! Collar} \email{Electronic address: collar@uchicago.edu}\affiliation{\UC}
\date{\today}

\begin{abstract}
Dark matter particles gravitationally bound to our galaxy should exhibit a characteristic speed distribution limited by their escape velocity at the position of the Earth ($v_{esc}\simeq$ 550 km/s). An ongoing search for anomalous cosmic rays at Earth, kinematically similar to cold dark matter, is described. The technique can discriminate between these and known slow-moving particles such as neutrons, would be sensitive to telltale signatures from presently unexplored candidates, and offers the possibility of identifying the mediating type of interaction (nuclear vs.\ electron recoils). Studies of background identification and abatement in a shallow underground site are presented. The expected reach of the method is discussed, and illustrated by obtaining the first limits for dark matter particles lighter than 100 MeV/c$^{2}$ interacting via nuclear recoils. 
\end{abstract}



\maketitle

\section{Introduction}

Over the three decades spanning since the first direct search for particle dark matter \cite{first}, a vast number of dedicated experimental techniques have been proposed and implemented, none of them returning an unambiguous evidence for dark matter interactions. With the exception of isolated searches for axion-like particles, workers in this field have predominantly concentrated on the detection of nuclear recoils induced by hypothetical Weakly Interacting Massive Particles (WIMPs) \cite{goodman,wasserman}, heavy neutral candidates with a cross-section for elastic scattering off nuclei comparable (but possibly much smaller) to weak-scale interactions. WIMPs are naturally generated by, for instance, supersymmetric extensions of the Standard Model \cite{wimps1,wimps2}. 

The combination of negative results from both direct searches and accelerator experiments has resulted in a progressive reduction of the available supersymmetric parameter space able to generate a cosmologically relevant WIMP \cite{susy1,susy2,susy3}. Perhaps as a reaction to this, the last few years have witnessed a surge of phenomenological interest in dark matter alternatives to a ``vanilla" medium-mass (10-1000 GeV) WIMP interacting through nuclear recoils. Two themes, oftentimes overlapping, can be discerned in this flurry of activity: candidates with lighter masses $m_{\chi}\lesssim$ few GeV \cite{model0,model00,model1,model2,model3,model4,model5,model6,model7,model8,model9,model10,model11,model12}, incapable or limited in their ability to produce signals above the energy threshold of present devices, and an examination of interaction mechanisms other than nuclear recoils \cite{int1,int2,int3,int4,int5,int6,int7,int8,int9,int10,int11,int12,int13,int14,int15,simpwind3}.  Unfortunately, the heavy investment of the experimental community into the next and possibly final generation of WIMP detectors has resulted in a certain inertia, with few (as of yet) searches being performed in response to this phenomenological prod.

An example of this quest for alternatives is a revival of interest \cite{simp1,simp2,simp3} in Strongly Interacting Massive Particles (SIMPs), candidates with a considerably larger interaction cross-section than WIMPs. SIMPs with masses in the MeV to GeV range and strong self-interactions are able to address astronomical puzzles hard to tackle with a standard dissipationless WIMP \cite{simpast1,simpast2,simpast3,simpast4,simpast5}. SIMPs may also play a role in the formation of ultrahigh energy cosmic rays \cite{rocky1,rocky2}. A recurring concept in this area is the existence of a window of unexplored dark matter phase space for $m_{\chi}\lesssim$ few GeV, and nuclear scattering cross-sections of O(1) micro-barn and above \cite{simpwind1,simpwind2,newfarrar}. A recent re-examination of this ``window of opportunity" has emphasized a broadening of possibilities whenever the nuclear recoil paradigm is abandoned \cite{simpwind3}. 

Two main reasons exist for the survival of this unexplored low-mass SIMP window. First, the mentioned limited ability of present detector technologies to sense interactions from slow-moving, low-mass particles: a 1 GeV/c$^{2}$ candidate traveling with a velocity of 300 km/s, typical of an object gravitationally bound to our galaxy, carries a mere 500 eV of kinetic energy. Regardless of the mechanism of interaction and technology involved, this does not leave much room for the generation of signals above detector threshold. This is particularly true of nuclear recoils, where the situation is aggravated by a limited maximum recoil energy, when imparted to a target heavier than the projectile, and by the so-called ``quenching factor" (a measure of the diminished ability of a recoiling ion to generate ionization or scintillation, when compared to an electron of the same energy). Second and related, the  overburden above deep underground laboratories housing WIMP experiments rapidly thermalizes low-mass particles with scattering cross-sections above $\sim10^{-6}$ barn \cite{int14,simpwind3,mod1,newea1,newea2}. A similar statement can be made about the Earth's atmosphere, for cross-sections $\sigma$/m$_{\chi} \gtrsim 10^{-2}$ barn/GeV \cite{balance}. As a result, WIMP searches performed at depth do not constrain candidates with sufficiently large cross-sections.

Not all are disadvantages when discussing SIMP direct detection. A highly-characteristic diurnal modulation in detection rate is expected from the shielding effect of the Earth on particles with sufficiently-large interaction cross-sections. The effect, first proposed in \cite{ourmod1,ourmod2,ourmod3}, arises from a preferred direction due to the Earth's motion through the galaxy, and its daily rotation, generating a latitude-dependent modulation (Fig.\ 1).  Sought for WIMPs in \cite{mod2}, this type of modulation has been recently revisited within the SIMP context \cite{simpwind3,mod1,mod3,mod4,mod5,mod6,mod7,mod8}.

\begin{figure}[!htbp]
\includegraphics[width=0.4\textwidth]{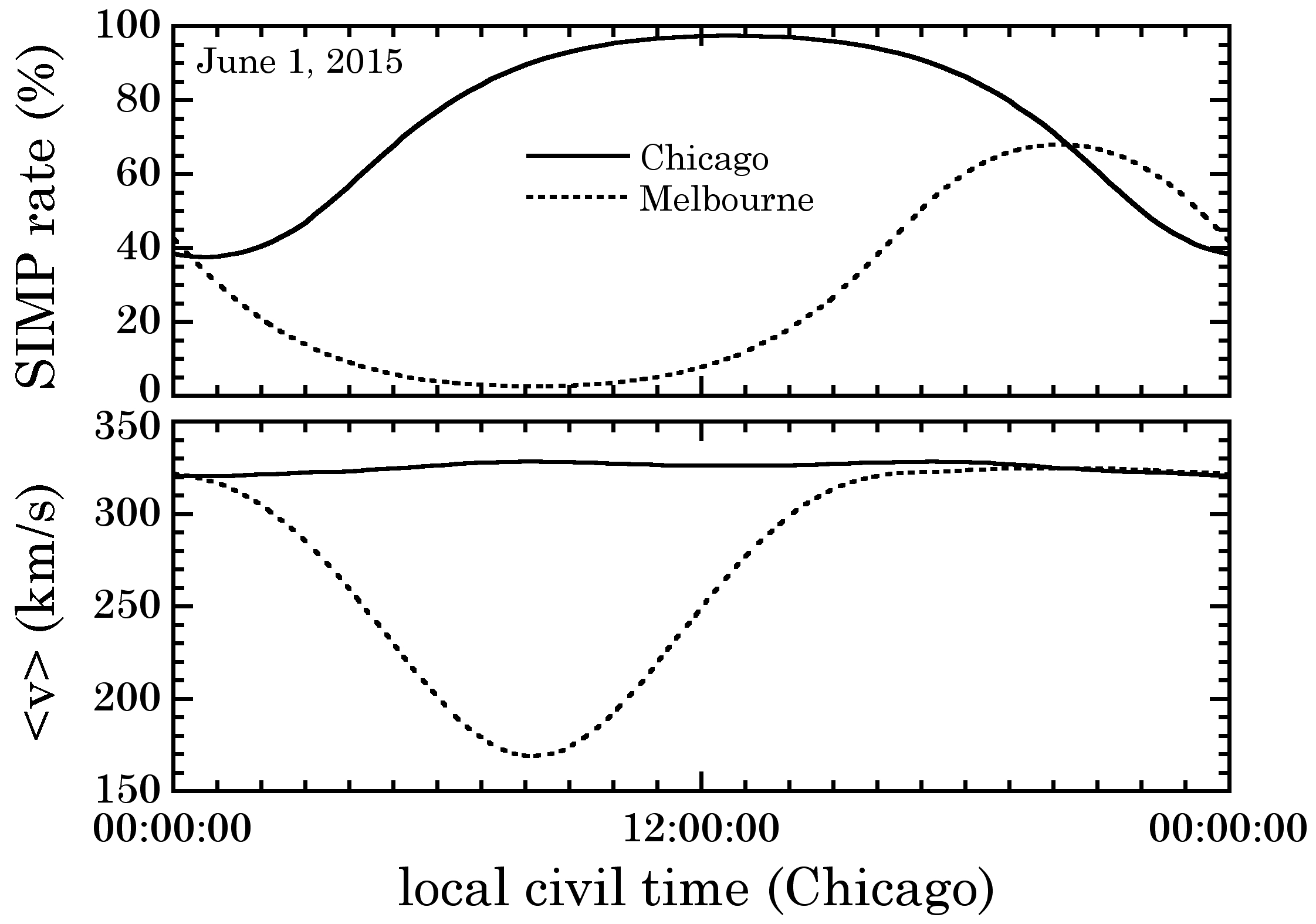}
\caption{\label{fig:lee} Diurnal modulation effect \cite{ourmod1,ourmod2,ourmod3,mod2,simpwind3,mod1,mod3,mod4,mod5,mod6,mod7,mod8} for the case of a SIMP blocked by the Earth, i.e., able to reach a shallow-depth detector only from above the horizontal. The calculation follows  \protect\cite{mythesis}, using astronomical subroutines adapted from \protect\cite{oxford}. The variation in interaction rate is relative to an unimpeded flux (100\%). The local time of the daily maximum,  asynchronous between locations, varies slowly through the year, creating a highly-characteristic signature. Searches measuring SIMP time-of-flight  may also observe its modulation, due to changes in SIMP average velocity (bottom panel).}
\end{figure}

\section{Experimental Approach}

This paper describes  technical aspects of an ongoing experimental program aiming to improve present sensitivity over a wide range of SIMP masses. An obvious requirement is to perform this type of search at much shallower depth than conventional WIMP experiments. This imposes a change in strategy, in view of the  higher levels of cosmic ray-associated backgrounds in such a site. One possibility is to rely on a delayed-coincidence technique, able to measure the characteristic SIMP time-of-flight between two detectors. While the coincidence requirement  impacts the reach of the search in interaction cross-section, it also results in a dramatic reduction in competing backgrounds. In Sec.\!\! V it is shown that once the dominant source of low-energy background (beta decays in scintillation detector windows) is identified and removed, a sensitivity to SIMP signals as infrequent as $10^{-2}$ events/kg-day is reachable. Sec.\ VI describes several detector configurations  tested, concluding that an exploration of virgin SIMP parameter space is possible.

\begin{figure}[!htbp]
\includegraphics[width=0.4\textwidth]{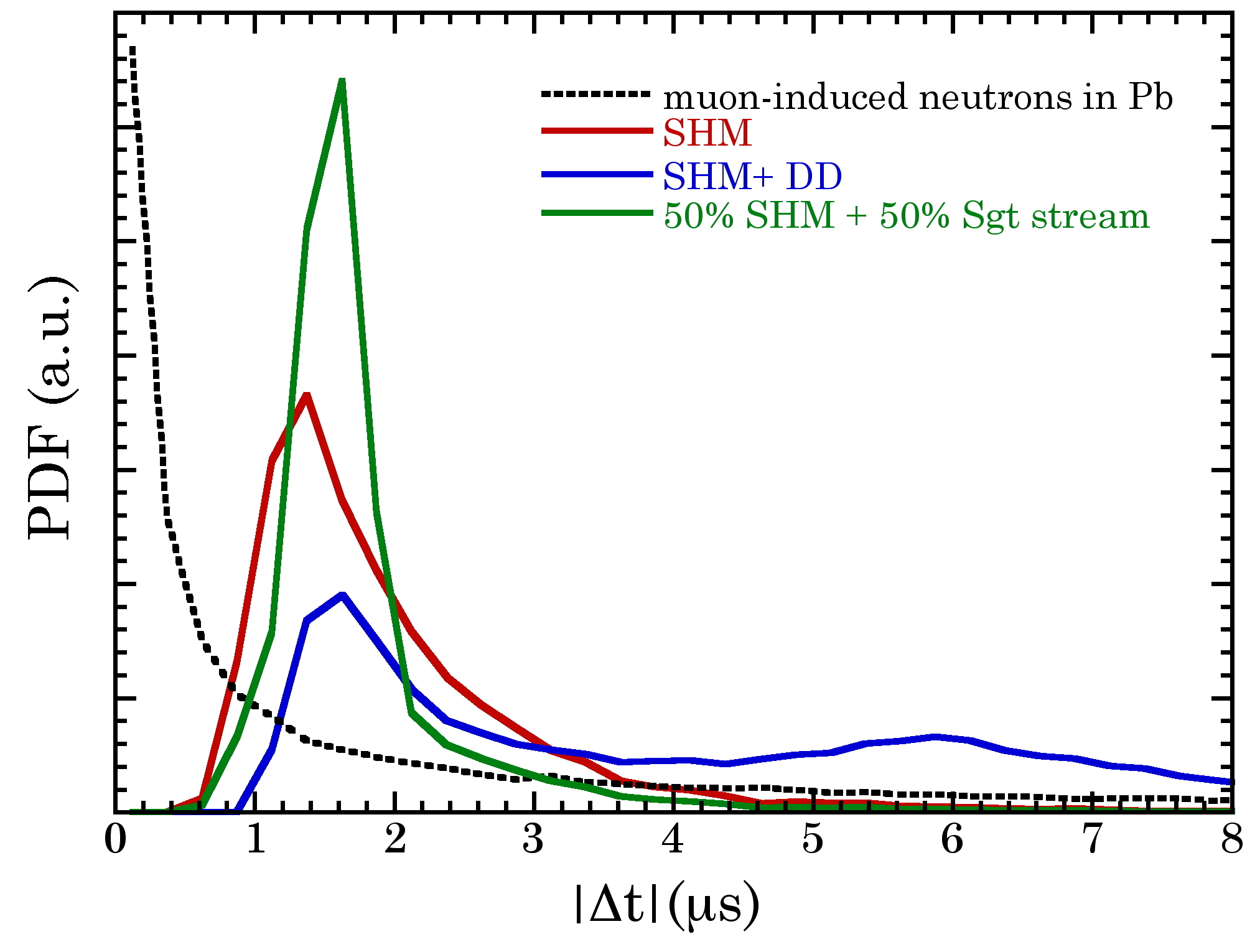}
\caption{\label{fig:lee} Normalized probability distributions of SIMP time-of-flight in present detectors, obtained by sampling different possible halo speed distributions in the Earth's reference frame (SHM stands for standard halo model, DD for dark disk, Sgt for Sagittarius) \protect\cite{halo1,halo2,halo3,halo4}. An ability to distinguish between halo models, and to discriminate against neutron-induced signals, is evident (see text). This calculation is only indicative, as it assumes isotropic dark matter trajectories, and ignores model-dependent SIMP energy loss in the first detector. }
\end{figure}

An implementation of the delayed-coincidence method is described in detail in the next section. It employs two hydrogenated liquid scintillator (LS) cells as detectors, able to provide information about the type of interaction creating the signals, down to very low energy. Fig. 2 displays SIMP time-of-flight between the cells (TOF, denoted by $\Delta$t), the measured quantity of most interest. This is specific of the present arrangement, where the travel distance between detectors is $\sim$60 cm. Overlapped on these TOF distributions is a calculated response to background neutrons, experimentally demonstrated in Sec. IV. An attractive property of this method is the good separation between SIMPs and neutrons, evident in the figure. A TOF of 2 $\mu$s over 60 cm, comparable to what is expected from a  SIMP gravitationally-bound to our galaxy, would imply a neutron kinetic energy of just 440 eV. A neutron this slow, even when transferring all of its energy to a proton recoil in a single scatter, has a small probability of generating scintillation in LS. While this same limitation would apply to an equally slow 1 GeV SIMP predominantly interacting via proton recoils, dark matter particles are potentially able to display other mechanisms of interaction \cite{int1,int2,int3,int4,int5,int6,int7,int8,int9,int10,int11,int12,int13,int14,int15,simpwind3}, not partaken by neutrons. Most importantly, values of $\mid\!\!\Delta$t$\mid$ smaller than $\sim1~ \mu$s are not possible  for dark matter particles, due to their characteristic halo velocity distribution, which is bound from above by the galactic escape velocity at the position of the Earth ($v_{esc}\simeq$ 550 km/s). In contrast to this, signals from environmental and muon-induced neutrons display a monotonically increasing rate with decreasing $\mid\!\!\Delta$t$\mid$, as larger kinetic energies result into readily detectable energetic proton recoils.

In principle, this method is able to return a number of SIMP signatures: {\it i}) a characteristic $\Delta$t distribution able to distinguish between galactic halo models (Fig.\ 2), {\it ii}) an asymmetry in sign of $\Delta$t, for SIMPs efficiently blocked by the Earth, {\it iii}) the mentioned diurnal modulation effect, and {\it iv}) for some proposed mechanisms \cite{int12}, a revealing difference in the type of interaction (nuclear recoils vs.\ electron recoils) involved in each detector.  In view of this promise, it is cautious to inspect the history of searches for slow cosmic-rays, looking for a possible redundancy. 

Early searches for fractionally charged and/or massive cosmic rays performed before the 1980s restricted their reach to a relativistic $\beta>0.1$ \cite{cr1,cr0,cr2,cr3,cr4}, in part due to assumptions derived from the geomagnetic rigidity cutoff. Numerous more recent searches for magnetic monopoles and nuclearites \cite{mm1} concentrated instead on the $\beta\sim10^{-3}$ characteristic of a galaxy-bound particle. However, the  large stopping power ($d$E/$d$x) expected from these dark matter candidates, up to a few thousand times that from a minimum-ionizing particle \cite{mm2}, led to detector thresholds set much higher than what is required for a comprehensive low-mass SIMP search \cite{mm3,mm4,mm5,mm6,mm7,mm8}. Taking for comparison the leading MACRO detector \cite{mm7,mm8}, the experimental arrangement described in Sec.\ III is set to trigger on scintillation light yields lower by a factor of seventy-five. Certain sensitive searches for SIMPs at sea-level \cite{mm5} depart from strong assumptions (continuous $d$E/$d$x via ionization, absence of any radiative losses) that can be relaxed in the present approach. Closer to this work in concept and motivation, are searches for SIMPs performed by the BPRS and DAMA collaborations using NaI(Tl) \cite{d1,d2}. Due to their siting at  a 3,400 m.w.e.\ overburden, they provide bounds for m$_{\chi}>10^{3}$ GeV only.

\section{Implementation}

The commercial organic liquid scintillator EJ-301 \cite{specs},  previously marketed under the denominations BC-501A and NE-213, is ideal for a SIMP search. This preference can be justified on the following grounds:

First, its large hydrogen content  ($4.8\times10^{22}$ H/cm$^{3}$, H:C ratio = 1.21) kinematically favors hard-to-reach dark matter particles of mass $\sim$1 GeV, whenever the interaction is mediated by nuclear recoils. In this case, up to the total kinetic energy of the particle can be transferred to a recoiling proton, in a single collision, improving the chances of generating a measurable scintillation signal. The material is also responsive to interactions via ionization or molecular excitation \cite{int1,int2,int3,int4,int5,int6,int7,int8,int9,int10,int11,int12,int13,int14,int15,simpwind3}. Similarly to other scintillators, these mechanisms require just a modest investment of energy in order to produce one scintillation photon (few tens of eV and few eV, respectively \cite{galunov}). This is an asset in a search for low-mass particles that makes no {\it a priori} assumptions about the mechanism of interaction. The scattering of sub-GeV dark matter off hydrogen atoms has been recently studied in depth, for several possible modes of interaction \cite{h1,h2}.

Second, EJ-301 is the scintillator of choice for pulse shape discrimination (PSD) between neutron-induced nuclear recoils (NRs) and gamma-induced electron recoils (ERs). In this material, a dissimilar ionization density left along the wake of a NR or ER results in large differences in the relative intensity of fast and slow scintillation decay components \cite{brooks}. Specifically for EJ-301, these components have decay constants of 5 ns (fast), 32 ns (medium), and $\sim$140 ns (long). ERs populate these channels in a 89:7:3 proportionality ratio, which becomes 56:25:19 for NRs  \cite{decays}. This PSD capability has been extensively characterized and exploited over the past few decades, but at energies larger than those of interest for a low-mass SIMP search. It is shown in Sec.\ IV that this PSD remains usable in the few photoelectron (PE) regime. This PSD capability may allow an eventual identification of the type of interaction mediating dark matter signal formation.

\begin{figure}[!htbp]
\includegraphics[width=0.43\textwidth]{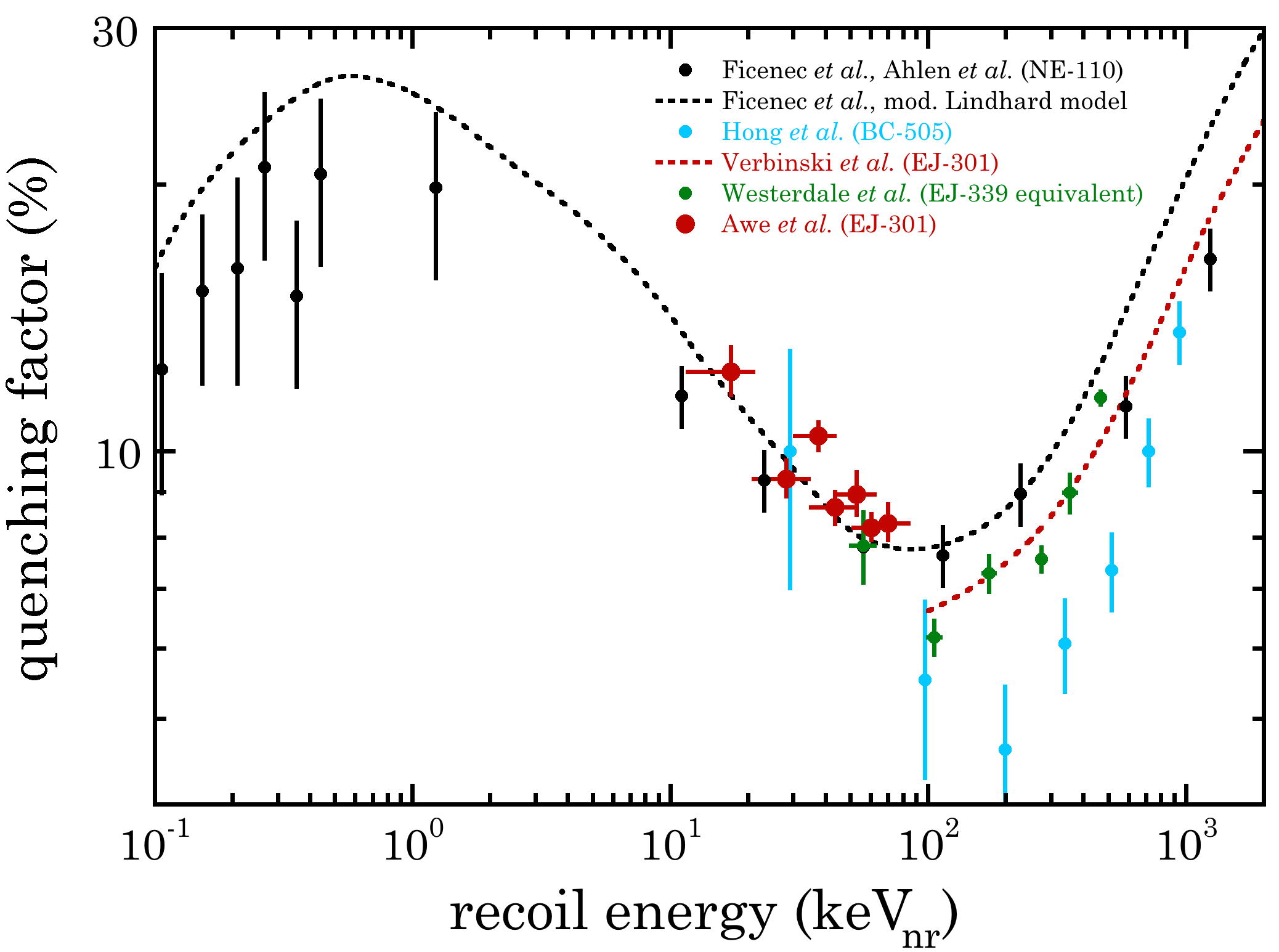}
\caption{\label{fig:lee} Low-energy quenching factor (QF) for proton recoils in organic scintillators \protect\cite{ahlen1,ahlen2,awe,ej301quenching,spooner,calaprice}. A black dotted line represents the modified Lindhard model proposed in \protect\cite{ahlen2}. Neutron simulations in Sec.\ IV and SIMP sensitivity expectations in Sec.\ VI adopt this model for EJ-301, and a power function fit to data from  \protect\cite{ej301quenching,spooner} for subdominant (QF $\sim$1\%) carbon recoils. Recent EJ-301 work by the author and collaborators (Awe {\it et al.}, \protect\cite{awe}) supports the presence of a large QF for sub-keV proton recoils, able to enhance low-mass SIMP sensitivity. }
\end{figure}

Third, keV and sub-keV proton recoils like those possibly expected from low-mass dark matter particle interactions, have a favorably large quenching factor in EJ-301 ($>$15\%,  Fig.\ 3), leading to detectable scintillation signals. This statement is based on calibrations performed using NE-110 \cite{ahlen1,ahlen2}, an organic plastic scintillator previously employed for monopole searches \cite{mm7,mm8}. A most recent calibration of EJ-301 response using monochromatic 245 keV neutrons (Awe {\it et al.}, Fig.\ 3) confirms a similar behavior is in place for all aromatic organic scintillators \cite{awe}. This is additionally supported by the neutron calibrations discussed in Sec.\ IV.   

Lastly, at 12,000 scintillation photons per MeVee (78\% of anthracene's emission) \cite{specs}, EJ-301 exhibits one of the highest light yields from liquid scintillators, facilitating the detection of weak signals. Its fast scintillation decay time makes it ideal for delayed coincidence measurements like those involved in some aspects of this search. As an additional bonus, organic liquid scintillators are typically quite radioclean, leading to lower internal backgrounds. This generally observed behavior is due to the small solubility of complex U and Th salts in aromatic solvents. 

The reminder of this section provides details about a delayed-coincidence system dedicated to this search, installed under the moderate overburden (6.25 m.w.e., a $\sim$60\% increase on atmospheric depth) available at the Laboratory for Astrophysics and Space Research (LASR, now PRC) of the University of Chicago. This location, originally designed as a low-background counting laboratory, features clean-room conditions under six feet of concrete. This limited overburden results in an order of magnitude reduction in environmental neutron flux, and the removal of soft and nucleonic cosmic ray components \cite{povinec}. This site provides an ideal compromise between background reduction, and a depth modest enough to allow for a putative SIMP population to reach the detectors, at least from the vertical. 

\begin{figure}[!htbp]
\includegraphics[width=0.4\textwidth]{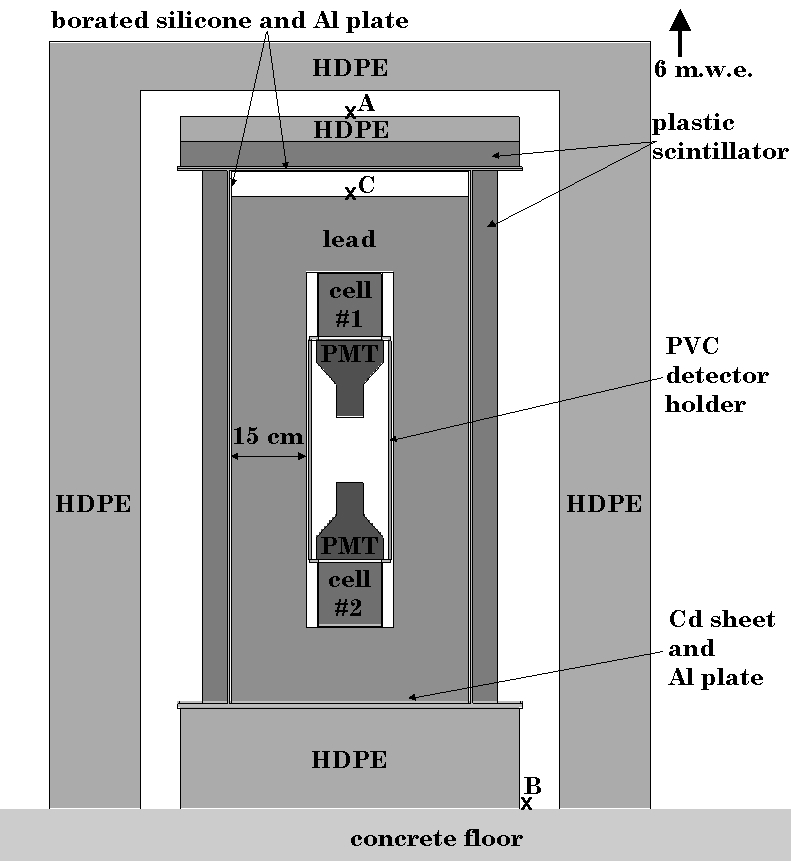}
\caption{\label{fig:lee} Shielding and positioning of detectors during this search, with all dimensions to scale. Lead thickness is provided as a reference. High-density polyethylene (HDPE) is used as neutron moderator, cadmium sheet and borated silicone acting as thermal neutron absorbers. An aluminum extrusion structure, not shown, is used for HDPE support. Three source positions used during calibrations (Sec. IV) are indicated by labelled crosses. }
\end{figure}

{\it Shielding and detectors:} the shield and initial detectors employed in this search were previously used to characterize neutron backgrounds at the Spallation Neutron Source (Oak Ridge National Laboratory), a study performed within the context of the COHERENT collaboration \cite{coherent}. A full description of this shield (Fig.\ 4) can be found in \cite{coherentnim,nicole}. An innermost 1"-thick ultralow-background lead liner was removed to accommodate the detectors, two commercial EJ-301 cells, each containing 1.5 l of active material housed within a 5" diameter, 5" long inner cell volume. Each cell is read out by one five-inch ET9390KB \cite{et} photomultiplier (PMT), favored for PSD applications \cite{luo}. The 58 cm center-to-center distance between the cells is a compromise between having a significant mutually-subtended solid angle, necessary to improve the probability of a SIMP traversing both, and enough separation to provide a SIMP time-of-flight of O(1)$\mu$s (Fig. 2), easy to resolve with the present system. 

Elements of this shield were rearranged to ensure the best possible horizontal symmetry of lead and neutron moderator materials (HDPE and plastic scintillator), so as to avoid a preferential vertical direction for incoming background radiations. The active veto against the hard (muonic) cosmic ray component is composed of seven 2"-thick plastic scintillator panels, each internally housing four 3/4" ET9078B PMTs \cite{et}. Each group of PMTs was gain-matched  to allow for a daisy-chained single output and power cable per panel \cite{nicole}. The internal mounting of the PMTs leads to a compact veto geometry and good light collection, resulting in an excellent separation of signals produced by environmental gammas and muon interactions \cite{nicole}. This separation allows for a modest (90 s$^{-1}$) triggering rate from the ensemble of the panels, in good agreement with muon flux expectations at this shallow depth. The six bevel-edged side panels conform to the hexagonal cross section of the shield, providing a muon veto efficiency in the 99.60\%-99.97\% range, depending on position within the enclosed volume \cite{nicole}. 

\begin{figure}[!htbp]
\includegraphics[width=0.47\textwidth]{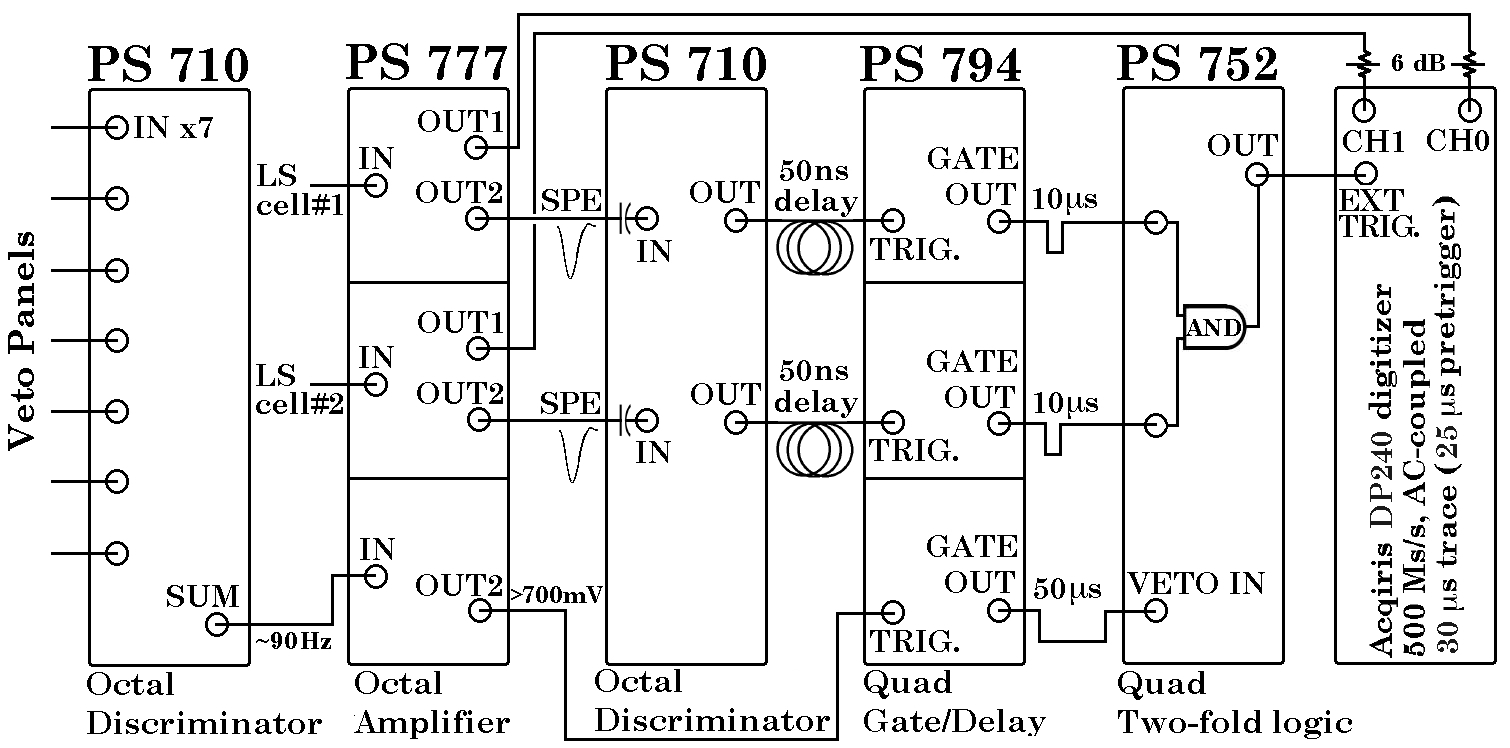}
\caption{\label{fig:lee} Schematic of the DAQ system (see text). The sum output from the first PS710 discriminator is proportional to the number of muon veto panels fired. It is amplified so that a single panel is sufficient to inhibit the PS752 logic module. Discriminator levels were adjusted to avoid excessive triggering on environmental gamma interactions with veto panels, while responding to muon-induced signals \protect\cite{nicole}. Inline 6 dB attenuators are used at the digitizer input to keep large (saturated) PMT signals from surpassing its maximum rating. }
\end{figure}

{\it Data acquisition (DAQ) system:} a trigger logic was developed using Nuclear Instrumentation Modules (NIM) from Phillips Scientific (PS). Its schematic is shown in Fig. 5. The bias to each of the two ET9390KB PMTs was adjusted to obtain a near-identical single photoelectron (SPE) response (Fig. 6, inset), corresponding to a PMT gain of 5$\times10^{6}$. The gain of the PS777 amplifier was set to its minimum value ($\times2$). Inline DC-blocking capacitors \cite{pomona} were used to improve the baseline stability at the inputs of the second PS710 discriminator in the figure, set to its minimum trigger level of -10 mV. The combination of PMT gain and discriminator level was selected to allow a majority of SPEs to generate PS710 outputs. The approximate discriminator level is indicated by a vertical dotted line in the inset of Fig. 6. 

A PS794 Gate/Delay generator was employed to issue prompt (11 ns maximum delay from trigger) 10 $\mu$s-wide logic gates on the arrival of a discriminator output signal. These gates are used as input to an ``AND" logic implemented via a PS752 module. The resulting PS752 output corresponds to the simultaneous appearance of SPE (or larger) signals in both LS cells within a 10 $\mu$s time window, regardless of which cell initiated the process. This output is used as a trigger to an Acqiris DP240 fast digitizer, with settings as indicated in Fig.\ 5. The digitizer trigger rate was a mere 4 s$^{-1}$, a result of efficient background shielding and modest dark count rate ($\sim$900 s$^{-1}$ at PS710 output) of ET9390KB PMTs.

\begin{figure}[!htbp]
\includegraphics[width=0.42\textwidth]{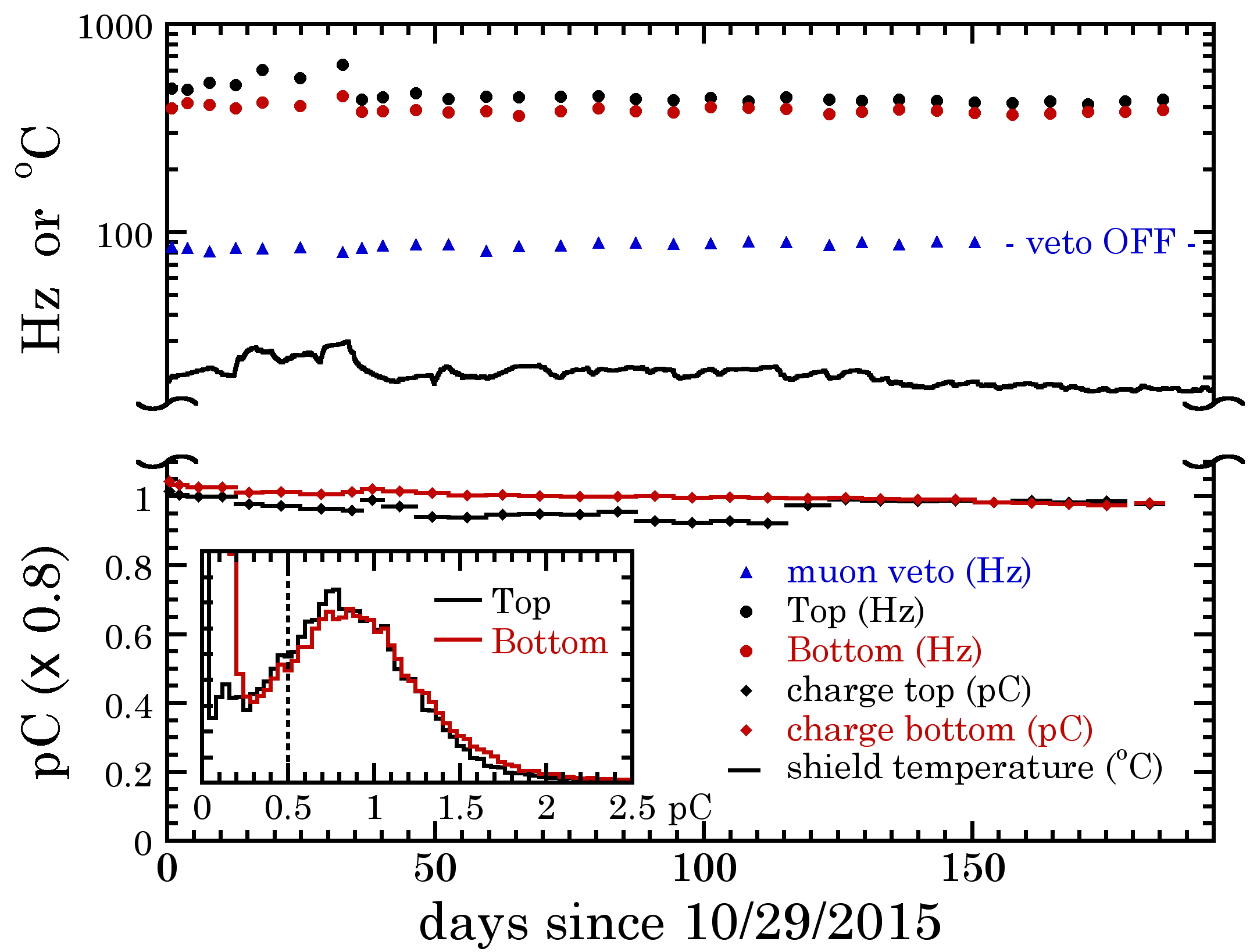}
\caption{\label{fig:lee} Long-term stability of LS cells (Top, Bottom) and veto signals. The SPE rate from ET9390KB PMTs was periodically measured at the PS794 output. Due to bunching, this rate is $\sim\!\times2$ smaller than at the corresponding PS710 outputs. The mean SPE charge (initial values shown in inset) was continuously monitored in the analysis of digitized traces. Veto rate is for the sum of all panels (individual panels were also periodically monitored). Room temperature was stable except for an early upward fluctuation during AC system adjustments, producing a small correlated SPE rate change.}
\end{figure}

The PS752 trigger to the digitizer was inhibited on the presence of a muon-veto signal. Special attention was paid to the length of signal cables involved, to ensure that the muon veto PS794 gate output correctly arrived to the PS752 unit in advance of related LS cell signals, also extending beyond their extinction. To this same end, 50 ns delays where added to LS cell PS710 outputs. A long 50 $\mu$s width for the muon veto PS794 output was selected to encompass the straggling of muon-induced neutrons, and to keep a majority of ET9390KB afterpulses following large muon-induced signals in LS from contributing to trigger generation. At the 90 s$^{-1}$ muon veto trigger rate, this 50 $\mu$s window generates a negligible 0.4\% dead time. 

A satisfactory long-term test of the stability of the full system is shown in Fig. 6.

{\it Data analysis:} Fig. 7 displays an example event, initiated by the top LS cell. The digitizer is programmed to trigger 25 $\mu$s into 30 $\mu$s-long traces. According to the stipulated  trigger logic, at least one of the two detector traces should display an onset of scintillation at t = 25 $\mu$s, with the other appearing anywhere in the preceding 10 $\mu$s. A LabVIEW analysis program implements a peak-finding algorithm able to extract this onset (t$_{0}$ in the figure), for each trace. The difference between t$_{0}$ values for cell \#1 (top in Fig. 4) and cell \#2 (bottom) is identified in what follows as the time-of-flight $\Delta$t. A negative (positive) value of $\Delta$t corresponds to an event initiated by cell \#1 (cell \#2). In terms of particle trajectories, SIMP-like cosmic rays efficiently shielded by the Earth would be expected to produce an excess of events at negative values of $\Delta$t only, obeying a distribution similar to those in Fig. 2. The peak-finding algorithm is observed to produce a negligible (0.2\%) failure rate in identifying at least one trace with t$_{0}$ = 25 $\mu$s. The algorithm implements a logic attempting to favor a multi-photoelectron scintillation flash over an isolated dark-current SPE when determining t$_{0}$. The moderate $\sim$900 s$^{-1}$ dark count rate from ET9390KB PMTs disfavors this possible source of t$_{0}$ misidentification.

\begin{figure}[!htbp]
\includegraphics[width=0.42\textwidth]{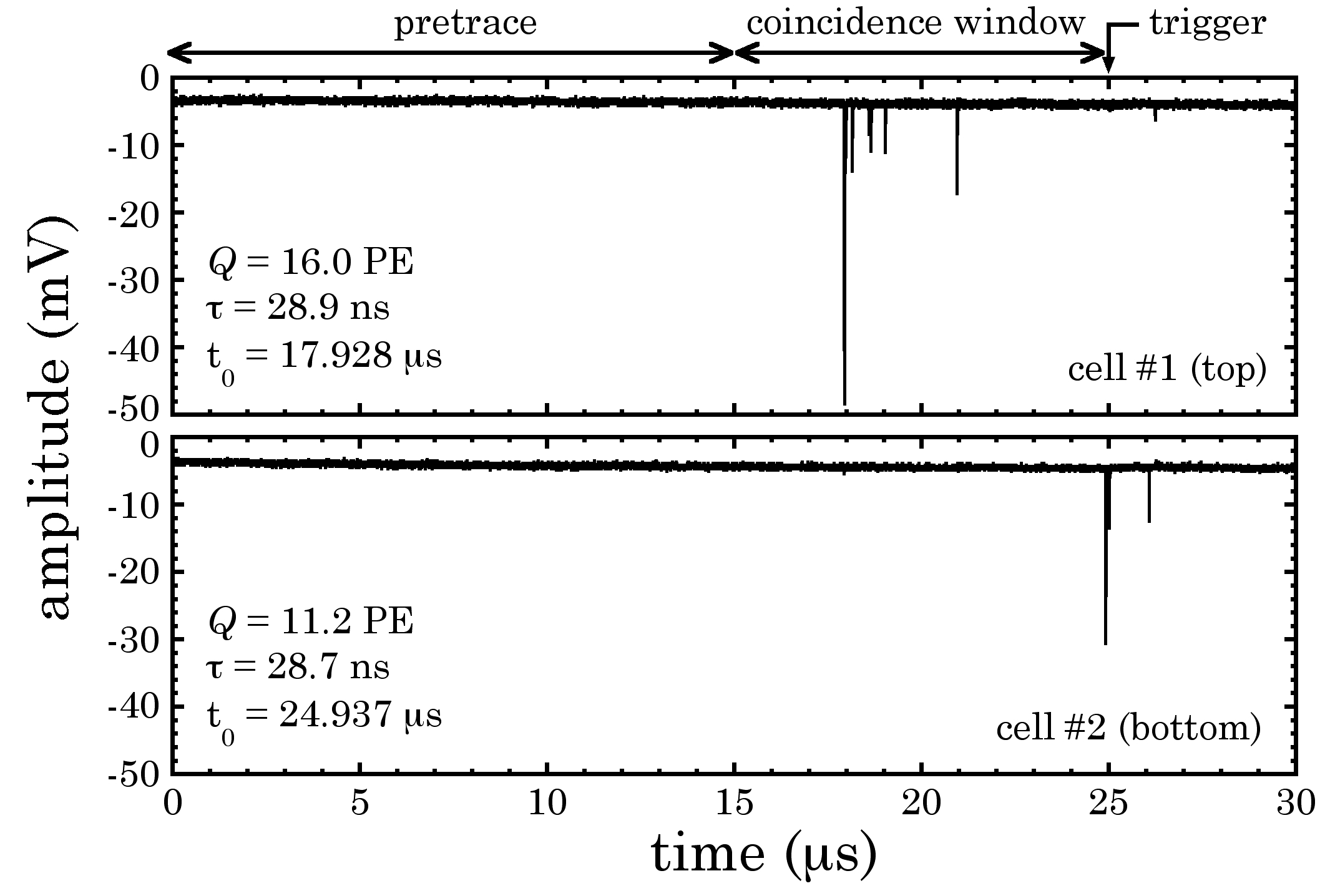}
\caption{\label{fig:lee} Example event ($\Delta$t$=\!\!-7.0 \mu$s) displaying the time regions and parameters described in the text. Afterpulses are visible following the primary scintillation, which is mostly contained within 300 ns after t$_{0}$ for EJ-301 \protect\cite{luo}. The digitizer is programmed to apply a $\sim$21 mV DC offset to the traces, in order to utilize the full 50 mV range available. Saturation of this range occurs for fast ($\tau<$ 20 ns) pulses with $Q\gtrsim$ 20 PE.}
\end{figure}

A 15 $\mu$s ``pretrace" preceding the 10 $\mu$s coincidence window is used to eliminate instances of PMT afterpulsing contributing to the generation of a trigger. PMT afterpulses are parasitic signals composed of single or few photoelectrons, following a true scintillation event. They originate in residual gas atoms or molecules within the PMT, backflowing towards the photocathode or dynodes after their ionization by the electronic cascade from the primary event \cite{aft1,aft2,aft3,aft4,aft5}. The bulk majority of PMT afterpulses appear a maximum of $\sim$15 $\mu$s after the primary, and hence the conservative length for this pretrace (specifically for the ET9390KB, this maximum delay has been reported to be $\sim$3 $\mu$s \cite{auger}). ``Slow" afterpulses appearing up to several hundreds of $\mu$s after a primary are comparatively rare, and limited to SPE generation \cite{aftslow1,aftslow2,aftslow3,aftslow4}. These can nevertheless generate a dominant background in some variants of a SIMP search, discussed in Sec.\ VI. A software cut against afterpulses can be applied on any events displaying scintillation in this pretrace region. Even for an aggressive cut removing events with just one SPE in either pretrace, event acceptance is high, at 96.4\%.

The analysis program returns a number of diagnostic parameters, including monitoring of SPE charge stability for each detector (Fig. 6), and of PMT afterpulsing as a fraction of primary scintillation. Following t$_{0}$ identification, the charge contained in each trace is integrated over the ensuing 300 ns, a period deemed optimal for PSD when using EJ-301 \cite{luo}. This charge can be expressed as an equivalent number of photoelectrons ($\it Q$ in Fig. 7). Several PSD parameters, including the decay time $\tau$ of the scintillation, are also extracted. These are described in the next section. Event time stamps are not available from this digitizer, but this information is extracted from the date of creation of the data spill containing the event. This provides timing information accurate within a few minutes, sufficient for an eventual search for the diurnal modulation effect discussed in Sec. I.

\section{Calibrations and Expected Backgrounds}

The detector assembly, DAQ system, and analysis pipeline described in the previous section were tested using neutron and gamma sources. These calibrations have multiple purposes: {\it i)} to show that EJ-301 provides usable PSD between NRs and ERs for signals comprised of just a few photoelectrons, an energy regime much lower than in conventional applications, {\it ii)} to test the applicability to EJ-301 of the modified Lindhard model for proton recoils (Fig.\ 3),  {\it iii)} to illustrate the ability to perform TOF and  directionality measurements on slow-moving particles (neutrons), and  {\it iv)} to confirm that neutrons and backgrounds associated to charged cosmic rays cannot mimick a slow-moving SIMP. Furthermore, the good agreement  between Monte Carlo predictions and calibration data validates the simulations of known backgrounds expected during a dark matter search. These findings are described in detail in this section.

An initial calibration using the 59.9 keV gamma emission from $^{241}$Am was used to establish an energy scale for the response to ERs. Good agreement was obtained with the 2.5 PE/keVee yield in \cite{decays}, once the modest non-proportionality of this scintillator \cite{nonlin} is included in the comparison (the notation ``ee" or ``nr" is used to denote ``electron equivalent" and ``nuclear recoil" in what follows). This same photoelectron yield can be calculated from the scintillation photon yield for EJ-301 (12,000 ph/MeVee \cite{specs}) and the quantum efficiency of ET9390KB PMTs integrated over the wavelengths of EJ-301 emission, indicating an excellent light collection efficiency for these cells. A small $\sim$15\% difference in light yield between top and bottom cells was noticed. This is a known effect from the positioning of an internal gas bubble \cite{bubble} used to avoid thermal expansion damage, and is neglected here.

\begin{figure}[!htbp]
\includegraphics[width=0.4\textwidth]{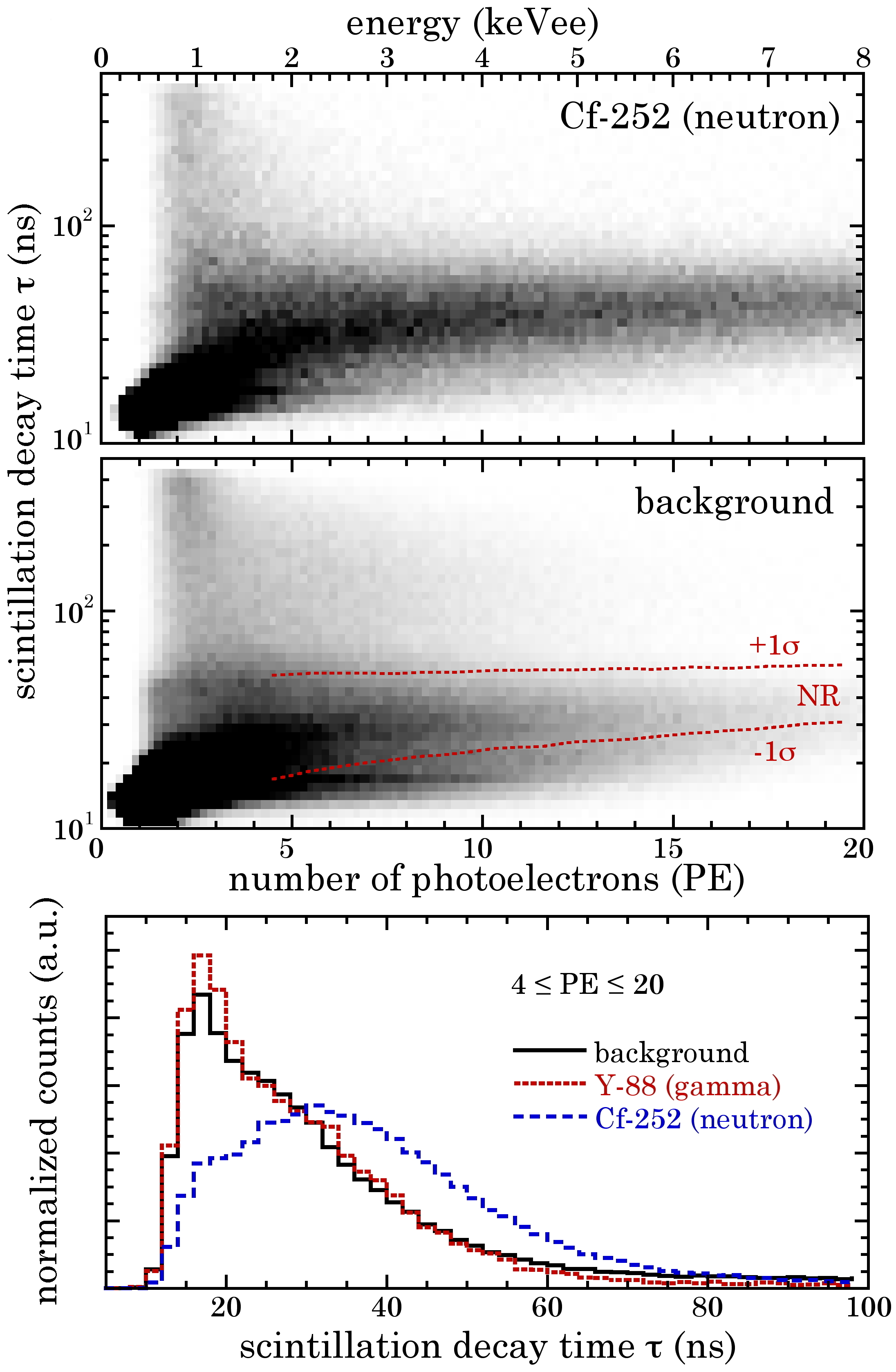}
\caption{\label{fig:lee} Top and middle panels: distribution of $\tau$ vs. energy for NRs from a $^{252}$Cf neutron source (2 h run), and background events (3 d run), for cell \#1. Fission gammas from $^{252}$Cf  are shielded by lead (position ``A" in Fig. 4). The $\pm1\sigma$ boundaries of the neutron-induced NR distribution (lognormal fit) are overlapped on background data, as a reference.  Bottom panel: normalized $\tau$ distributions for these events in the 4 PE $<Q<$ 20 PE energy region, including ERs from an $^{88}$Y gamma source (position ``C").  Penetrating high-energy gammas from this intense (1 mCi) source produced an increase over background trigger rate by a factor of seven. Background events are dominantly ER-like, as expected. }
\end{figure}

An abundant literature exists on analysis methods for ER/NR discrimination using EJ-301 and other PSD scintillators. This discrimination is based on the already mentioned differences in scintillation decay time from both types of interaction \cite{brooks}. The widely used integrated rise-time (IRT) method \cite{luo,ronchi} consists of digitally constructing an integrated scintillation curve for each event, finding the time difference between its crossing of two levels, defined as fractions of its maximum amplitude. For ET9390KB PMTs, optimized levels offering the best PSD are 10\% and 92\% of this maximum \cite{luo}. Gamma and neutron interactions depositing energies above few tens of keVee are sorted into two well-separated populations in this time difference, with a tendency for them to merge as the energy becomes smaller \cite{luo,ronchi}.

A simplified version of this method is illustrated in Fig. 8, where the parameter used for PSD is the scintillation decay time $\tau=t_{0-50}/$ln(2), fitted by a single exponential. Here $t_{0-50}$ is the time difference between the 0\% (t$_{0}$, onset of scintillation) and 50\% level crossings. As can be observed in the figure, NRs induced by a $^{252}$Cf neutron source and ERs from an $^{88}$Y gamma source display rather dissimilar distributions in $\tau$, even for pulses containing just a small number of PEs. An advantage of this parameter is that it can be directly compared to the decay constants characteristic of EJ-301 scintillation (Sec. III, \cite{decays}). As expected, NRs display a marked shift in $<\!\!\tau\!\!>$ towards the medium component scintillation decay constant of 32 ns, departing from the fast component decay constant of 5 ns that dominates for ERs (a rise time of 13 ns intrinsic to ET9390KB PMT multi-electron signals \cite{9390} shifts this fitted ER decay time to $\sim$18 ns). 

While the NR/ER separation presently observed at few PE is not optimal for an unambiguous event-by-event discrimination, it can suffice for statistical identification of the interaction mode mediating a low-mass dark matter signal, if sufficiently above backgrounds.  To this author's knowledge, this is a first demonstration of potentially usable PSD in the few PE regime for organic liquid scintillators. With a lowered PMT gain, as in the ongoing high-mass SIMP search mentioned in Sec.\ VI, the present system displays the usual excellent NR/ER separation characteristic of EJ-301 (Fig.\ 9, \cite{decays,luo,ronchi}).

\begin{figure}[!htbp]
\includegraphics[width=0.4\textwidth]{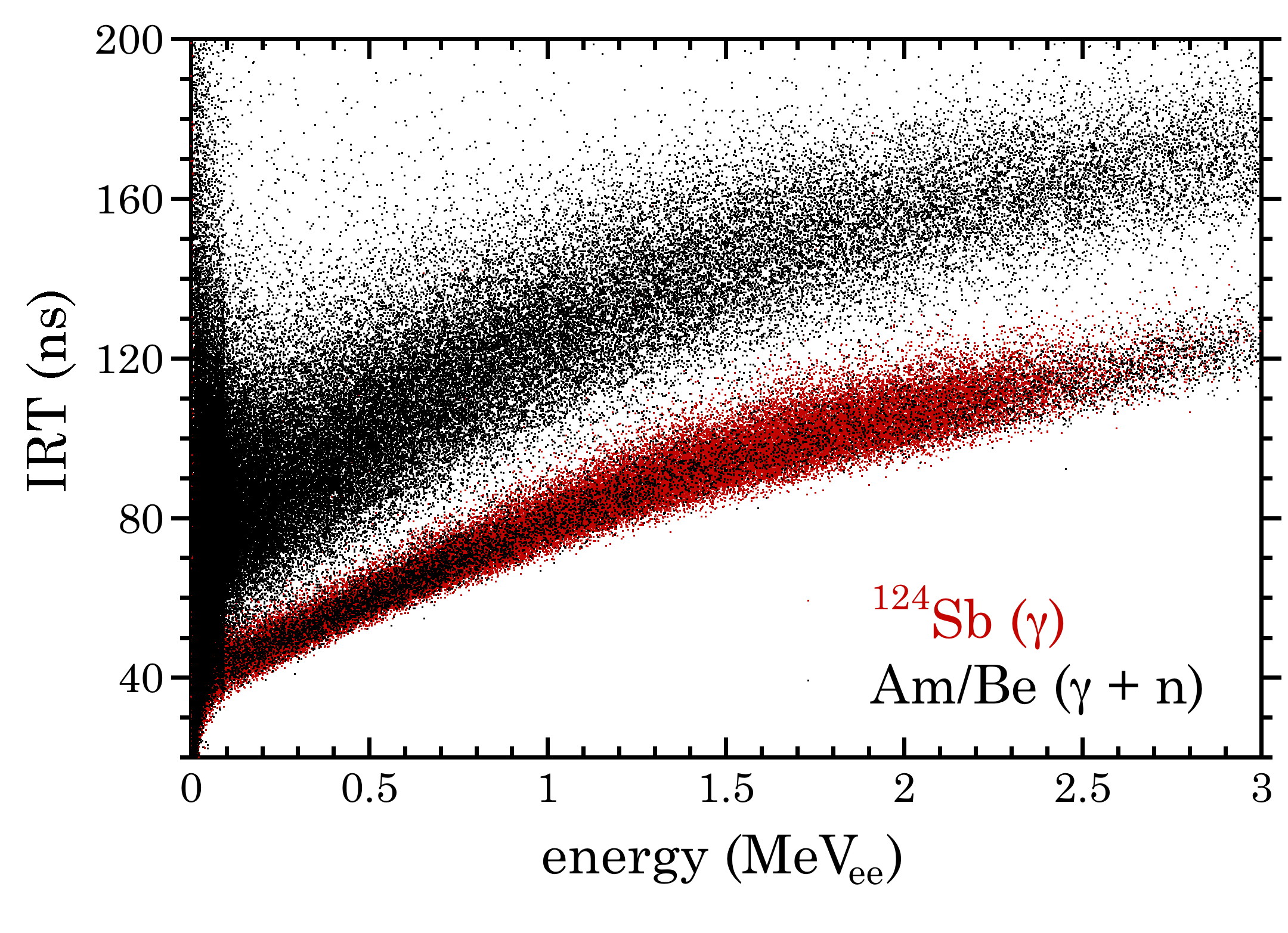}
\caption{\label{fig:lee} Separation between gamma-induced ERs and neutron-induced NRs with a reduced PMT bias, used during a high-mass SIMP search run (Sec.\ VI). The excellent PSD characteristic of EJ-301 \protect\cite{decays,luo,ronchi} is observed from few tens of keV up to few tens of MeV. }
\end{figure}

$^{252}$Cf neutron calibrations were performed with the source located at two different positions within the shield. The muon veto was switched off during these, to avoid an excessive dead time from its response to the source. For position ``A" (Fig.\ 4), the delay $\Delta$t between cell signals generated by the same neutron should have a predominantly negative sign, due to the proximity of the source to the top cell. The converse is expected from position ``B".  This effect is visible in the top panels of Fig.\ 10, demonstrating the ability of this system to measure the TOF and incoming direction of slow-moving particles. 

\begin{figure}[!htbp]
\includegraphics[width=0.4\textwidth]{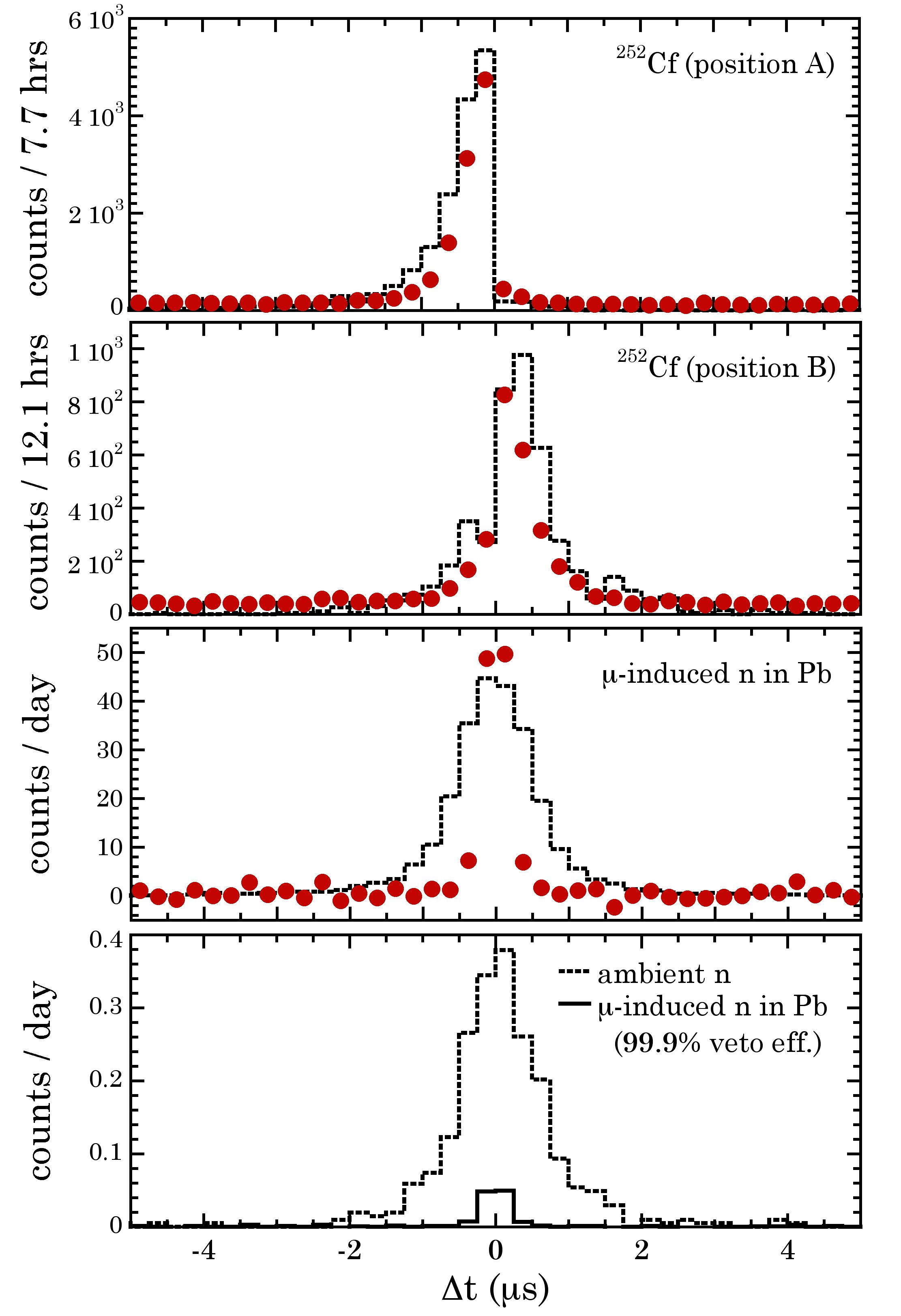}
\caption{\label{fig:lee} Simulated (histogram) and measured (data point) response to neutron calibration sources for coincident signals within the interval 3 PE$<Q<$14 PE ($\sim$1.2-5.6 keVee) and 10 ns $\!<\!\tau\!<\!$ 150 ns in both cells, together with expectations from known backgrounds (bottom panel). A negative $\Delta$t corresponds to a particle first striking cell \#1 (top), then \#2 (bottom), and conversely for $\Delta$t $>0$. Random coincidences uniformly affecting $\Delta$t were not simulated. Data points in the third panel are the normalized residual from 33 d of operation with the veto switched off, taking 150 d of veto-on operation as the background reference.}
\end{figure}

MCNP-Polimi \cite{polimi} was used to simulate the NR energies deposited in each cell, as well as the time difference between signals, employing a detailed geometry of the assembly (Fig.\ 4). The yield of the source ($8.0\!\times\!10^{5} ~\pm$ 10\% n/s) was used as an input. This yield originates in manufacturer specifications, independently confirmed in \cite{fustin}. Emitted neutron energies were sampled from a Watt spontaneous fission spectrum specific of $^{252}$Cf. Neutron-induced gammas from inelastic scattering or capture were included in the simulation. Following neutron transport, each simulated proton recoil energy in the cells was converted into an electron-equivalent deposition via the modified Lindhard quenching factor in Fig.\ 3. This was in turn translated into a PE yield through the 2.5 PE/keVee from $^{241}$Am calibrations. This yield is considered a central value around which Poisson fluctuations are applied. Subdominant carbon recoils were similarly processed (Fig.\ 3). Attention was paid to the timing of multi-site interactions within a cell in correctly determining the simulated $\Delta$t. Fig.\ 10 overlaps the simulated response over calibration data. They are in good agreement, even in the absence of any free parameters. 

To further test the adequacy of the modified Lindhard model, simulations were repeated with an alternative proton recoil quenching factor falling rapidly towards zero at $\sim$40 keVnr, i.e., missing the rise at lower energy visible in Fig.\ 3. This results in a similar $\Delta$t distribution, however overpredicting the measured rate in the 3-14 PE energy range by a factor of two. Considering that TOF between cells for a 100 (10) keV neutron is 0.4 (1.2) $\mu$s, and that the scattering of a neutron can lead to a total energy transfer to a proton, present $^{252}$Cf calibrations explore the response to proton recoils down to at least a few keVnr, complementing other measurements shown in Fig.\ 3. The absence of excess neutron coincidences beyond $|\Delta$t$|\gtrsim1~ \mu$s in the data points of Fig.\ 10 confirms the good separation expected between neutron backgrounds and galaxy-bound dark matter particles, illustrated by Fig. 2. 

An additional form of calibration involving harder neutrons is possible by temporarily switching off the muon veto. Cosmic ray muons traversing the lead shield are expected to generate a broad spectrum of neutron energies via capture ($\mu^{-}\!+p\rightarrow n+\nu_{\mu}$), photonuclear, and photo-fission reactions. These generate a dominant source of evaporation neutrons with energies below 4.5 MeV, and a smaller component of direct neutrons reaching out to higher energies, with exponentially decreasing spectrum \cite{dasilva}. Experimental data and simulations for this source are shown in the third panel of Fig. 10. The $\mu$-induced neutron production in the 2.09 ton lead shield was calculated by first fitting existing ($\mu,n$) reaction rate data for iron, collected in the range 20-200 m.w.e.\ \cite{russian1,russian2,russian3}. The obtained fit is $R=1.11\!\times\!10^{-4}\!\cdot d^{-1.429}$, where $R$ is in neutrons per g of Fe per s, and $d$ is the overburden in m.w.e. This is extrapolated to $d=$ 6.25 m.w.e., for the LASR laboratory. This neutron yield in iron is then converted to a lead equivalent through a scaling factor $A^{\beta}$, where A is the atomic weight of the target material and $\beta = 0.76\pm0.01$ \cite{scale,scale2}. The resulting muon-induced neutron generation in the lead shield is $3.95\times10^{6}$ n/day. This is used as an input to the simulation, with the assumptions of homogeneous production, and isotropic emission. The simulated spectrum of direct and evaporated neutron energies, extending out to 20 MeV, was generated following the prescription in \cite{dasilva}. Photoelectron yields and values of $\Delta$t are extracted as above. Fair agreement is once more observed between simulation and data, without the need for any free parameters. The measured distribution of $\Delta$t is however narrower than predicted by the simulation, possibly pointing at a larger proportion of energetic (i.e., faster) neutrons than what is suggested in \cite{dasilva} (that reference assumes a neutron emission solely from $\mu$-capture, while only half of the $\mu$-induced neutrons produced at shallow depth come from this process \cite{coco}). 

Motivated by the good match between simulations and calibration data, it is possible to predict the impact of  known neutron backgrounds on a dark matter search at this site. Two main sources are expected: first, cosmic-secondary and environmental neutrons from ($\alpha$,n), fission and ($\mu$,n) reactions in laboratory walls, floor and ceiling \cite{heusser}. Second, unvetoed $\mu$-induced neutrons in the lead shield. The flux and spectrum from the first source is known for this laboratory, albeit only over four coarse energy bins. This information was obtained by deconvolving measurements performed with a $^{3}$He counter surrounded by a number of moderator and absorber arrangements \cite{robynne}. For neutron energies within 0.5 eV-100 keV and  100 keV-10 MeV, most relevant to our concern, the respective measured fluxes were $5.6\times10^{-4}$ n/cm$^{2}$s and $6.9\times10^{-4}$ n/cm$^{2}$s, both an order of magnitude lower than at ground level. This rough spectroscopic information is used as input to a simulation assuming isotropic neutron trajectories originating outside of the HPDE shielding. The resulting expectation, shown in the bottom panel of Fig.\ 10, should be considered a conservative upper limit, since a flat distribution of neutron energies within the input energy bins was assumed, whereas a more realistic approach would use a monotonically-decreasing spectrum with increasing energy \cite{ziegler}. Regardless, the predicted background event rate with $|\Delta$t$|\gtrsim1~ \mu$s, where dark matter signals would accumulate, is very small. The same can be said of unvetoed $\mu$-induced neutrons, when the measured veto efficiency of $\sim$99.9\% is adopted.

To finish this section, a word of caution must be expressed about {\it unknown} backgrounds affecting SIMP searches. While much knowledge about low-energy processes has been accumulated through three decades of WIMP searches, these have been performed at large overburdens, minimizing effects associated to cosmic rays. Before embracing a SIMP-like excess in this style of search, characteristically restricted to $\Delta$t $\lesssim-1~ \mu$s, a careful consideration would have to be given to any cosmic ray-induced reactions potentially able to generate such a time-asymmetric signal. The same can be said of the diurnal modulation effect discussed in Sec.\ I, and known subtle day-night variations in cosmic ray flux \cite{day,day2}. An immediate example of these concerns involves muon stopping and capture, given the asymmetry expected from Earth shielding of these cosmic rays. Sub-keV muons nearing their stopping, slow enough to have a TOF between cells of O(1) $\mu$s, cannot generate signals in both cells due to their large $\sim$20 MeV/cm stopping power in hydrogenated liquid scintillator \cite{muon1,muon2,muon3}. However, upon decay or capture following traversal of the top cell, the engendered electron or neutron can be responsible for a delayed signal in the bottom cell. These scenarios, when arising from any charged cosmic ray able to trigger the veto, are nevertheless already highly constrained by the magnitude, symmetry, and narrow distribution of data points in the third panel of Fig.\ 10, and the high efficiency of the muon veto employed.

\section{ internal background abatement}

The system described above was continuously operated for 183 days, displaying good stability (Fig.\ 6). Its muon veto was switched off during the final 33 days of this initial test period, in order to obtain information about low-energy backgrounds associated to charged cosmic-rays (third panel in Fig. 10). 

Fig.\ 11 displays the time-of-flight distribution of events depositing a charge 4 PE $< Q <$ 14 PE ($\sim$1.6-5.6 keVee), and residing in the interval 10 ns $< \tau <$ 70 ns, for both detectors. This range of $\tau$ was chosen to include the majority of radiation-induced events (Fig.\ 8). The charge range can be extended down to 3 PE at the expense of just a small increase in background rate. Similarly, it can be extended up to $\sim$20 PE before saturation of the digitizer range occurs (Fig.\ 7). However, below $\sim$2.5 PE the background rate grows very fast, due to most triggers resulting from random coincidences between uncorrelated dark-current SPEs in each PMT (2.5 PE corresponds to 2 pC in the SPE charge distributions of Fig.\ 6). This affects the sensitivity of a SIMP search. This limitation could be  relaxed by monitoring LS cells with two PMTs each, requesting triple PMT coincidences as a trigger condition. Dynode glow \cite{dynode1,dynode2,dynode3}, a phenomenon able to produce correlated low-energy signals in PMTs facing each other, should however be kept in mind when considering such possible improvements. Use of high quantum efficiency super-bialkali PMTs such as the Hamamatsu R877-100 can lead to additional gains in minimum detectable energy and low-mass SIMP sensitivity. However, this author has gathered observations similar to those in \cite{hamathesis}: the smaller maximum gain and larger dark count rate of this alternative PMT renders it much harder to implement than ET9390KB's, for this application. A third possible avenue for improvement, discussed in Sec.\ VI, is to simply reduce the operating temperature of PMTs, and their dark current with it.

\begin{figure}[!htbp]
\includegraphics[width=0.5\textwidth]{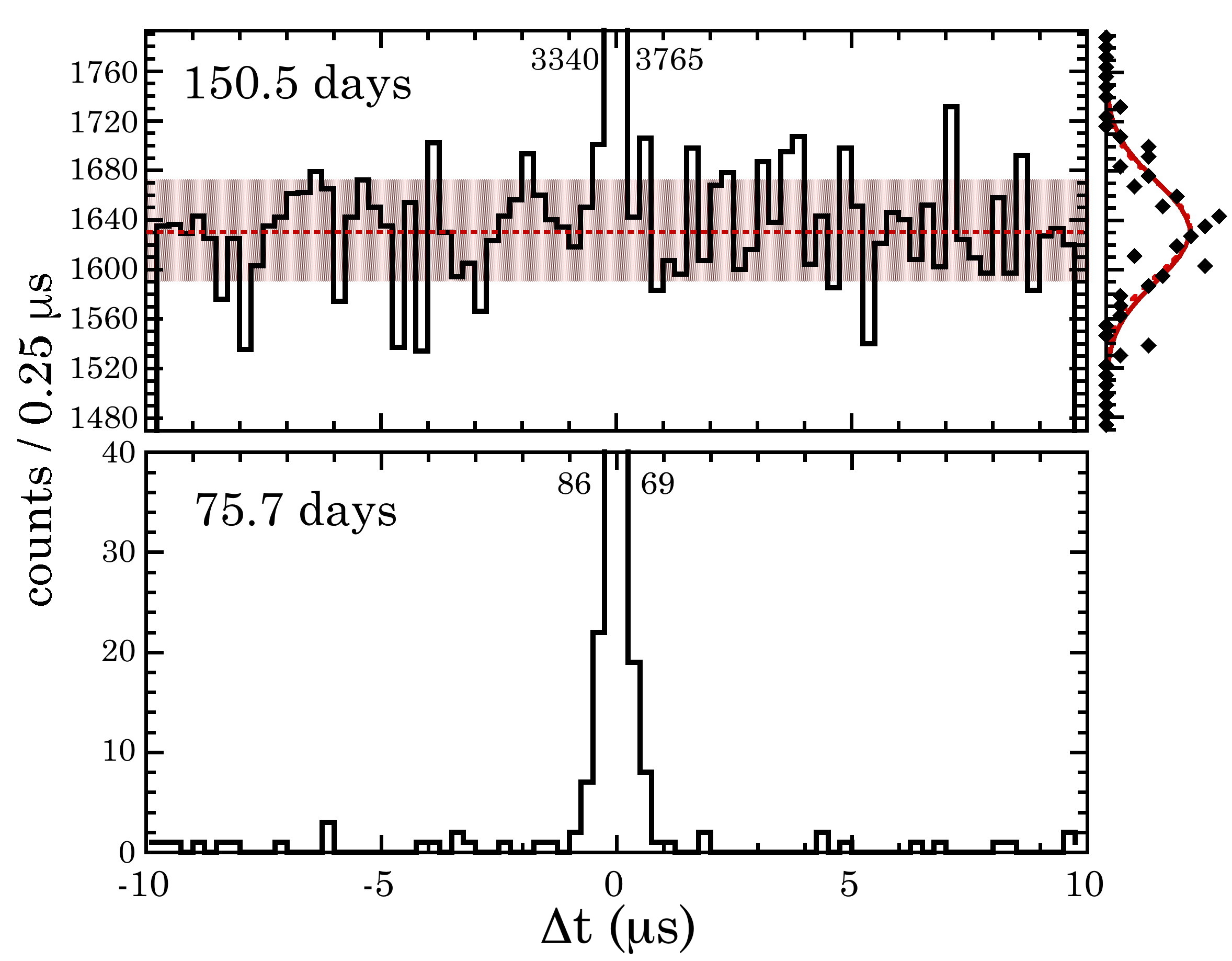}
\caption{\label{fig:lee} TOF distribution of events producing 4 PE $< Q <$ 14 PE ($\sim$1.6-5.6 keVee) in each detector, before (top) and following (bottom) background-reduction measures (an improvement by a factor of $\sim$1,650 for $|\Delta$t$|>1\mu$s ). A red dotted line indicates the  background mean, and a red band its $\pm1\sigma$ dispersion.  A side figure shows the Gaussian fit and Gaussian expectation of histogrammed bins, essentially indistinguishable from each other. The uniformity across $\Delta$t during initial  runs confirms an absence of systematics in t$_{0}$ determination. The content of bins out of range is shown. }
\end{figure}

Excluding the two central bins in the top panel of Fig.\ 11, dominated by prompt coincidences from gamma interactions, the event rate obtained was of approximately 35 coincidences per hour. This initially elevated background rate had one unexpected beneficial outcome: demonstrating the absence of any obvious artifacts in hardware or analysis, as evidenced by a good uniformity in event distribution across $\Delta$t, visible in the figure. Assuming the same backgrounds are affecting both cells, each would have to sustain an interaction rate of approximately 22 s$^{-1}$ in this few PE energy range, in order to generate the observed random coincidences. Previous experience with low-background detectors of similar mass, within this shield and in this same location, indicates that background rates three orders of magnitude lower than this can be obtained \cite{coherentnim}. Simulations using as input the known content in radioactive U, Th, and K in detector components (ET9390KB PMTs, Pyrex LS cell windows, Aluminum LS cell body, reflective paint) confirmed that these sources fall short by a factor of several hundred, in providing an interaction rate per cell that would explain these early observations. 

Inspection of the spectral shape of coincident events revealed a plateau above $\sim$3 PE, extending out to 10-12 PE, and rapidly dropping beyond, for both cells. This is characteristic of Cerenkov light emission in PMT and cell glass windows from beta emitters within \cite{cerenkov1,cerenkov2}. An estimate of beta emission rate from impurities in ET9390KB borosilicate windows can be extracted from typical values provided by the manufacturer (window mass 163 g, with $\sim$4,200 ppm K, $\sim$420 ppb Th, $\sim$380 ppb U). Assuming equilibrium in the radioactive chains, this results in $\sim$24 betas/s, per window. For Pyrex LS cell windows (197 g, typically $\sim$0.008\% natural K, $\sim$370 ppb Th, $\sim$580 ppb U), this results in an additional beta emission of $\sim$10 s$^{-1}$, per detector. The estimated combined rate of Cerenkov emission is therefore in fair agreement with the rate of 22 s$^{-1}$ necessary to generate the observed random coincidences, confirming this process as a plausible dominant contributor to the low-energy background. 

Use of synthetic fused silica in both PMT and LS cell windows was estimated to lead to a drop by a factor 30 in beta emission, resulting in a potential three orders of magnitude reduction in coincident background rate. Prompted by this, ET9390KB PMTs were replaced by ET9390QKA equivalents, identical except for their fused silica windows. These new PMTs were specially selected for photon-counting, leading to a reduction in dark count rate, compared to the original units. New LS cells where built using Corning 7980 synthetic silica windows ($<$5 ppb K, $<$0.5 ppb Th, $<$1 ppb U). Taking advantage of this opportunity, several other improvements were made to these cells:  their cans were built in Outokumpu 316L stainless steel ($\sim$10 ppm natural K, $<$10 ppb Th, $<$10 ppb U \cite{Outokumpu}), titanium oxide and sodium silicate solution used in reflective paint formulation were screened and selected for the highest chemical purity available, and commercially-available low-background EJ-301L scintillator was used. These additional measures can be considered a realistic best effort at reducing internal backgrounds in a shallow-depth experiment, where cosmic-ray sources are expected to eventually limit progress.  

Data from 75 days following these detector and PMT upgrades are shown in the bottom panel of Fig.\ 11. An overall background reduction by a factor $\sim$1,650 is observed, confirming that Cerenkov light from beta-emitting impurities was indeed the dominant low-energy source. A drop in prompt coincidence rate by $\sim45$ (two central $\Delta$t bins), demonstrates a significant decrease in gamma flux internal to the shield, also achieved through these detector and PMT modifications. The irreducible prompt coincidence  rate is just a factor of two larger than the simulated contribution from environmental neutron backgrounds (Fig.\ 10, lower panel).  The modest dark count rate ($\sim$200 s$^{-1}$) in the selected ET9390QKA PMTs is also beneficial to the background rate, when further reducing detector threshold. 

\section{ongoing experimentation and expected sensitivity}

This final section provides a brief description of present data-taking and the resulting expected SIMP sensitivity. Final results from the activities delineated below are in preparation \cite{common}. 

An ongoing delayed-coincidence search for high-mass SIMPs concentrates on the ``inflaton" mass scale around $10^{10}-10^{12}$ GeV. This mass range is of cosmological interest \cite{infla1,infla2,infla3,infla4}. The possible involvement of such dark matter candidates in the generation of ultra-high energy cosmic rays has been widely discussed \cite{rocky1,rocky2,uhe1,uhe2,uhe3}. To access the relevant SIMP phase space, PMT gain is relaxed in comparison to the background and characterization studies described in the previous sections. Present settings allow to identify energy depositions in the range 10 keVee - 30 MeVee, following corrections to account for DAQ range and PMT current saturation. In these conditions, an excellent separation between ER- and NR-mediated interactions is available (Fig.\ 9). A preliminary analysis indicates that a sensitivity to heavy SIMP-nucleus interactions with rate $<$0.1 day$^{-1}$ is achievable. This targeted search, motivated in \cite{jandc}, provides access to a presently weakly-constrained region of heavy SIMP phase space (cross-section vs.\ mass). The generality of previous experimental limits \cite{laurab,kava} can also be improved, by relaxing the assumptions of NR-mediated interactions and muon-veto anti-coincidence made in those.

\begin{figure}[!htbp]
\includegraphics[width=0.43\textwidth]{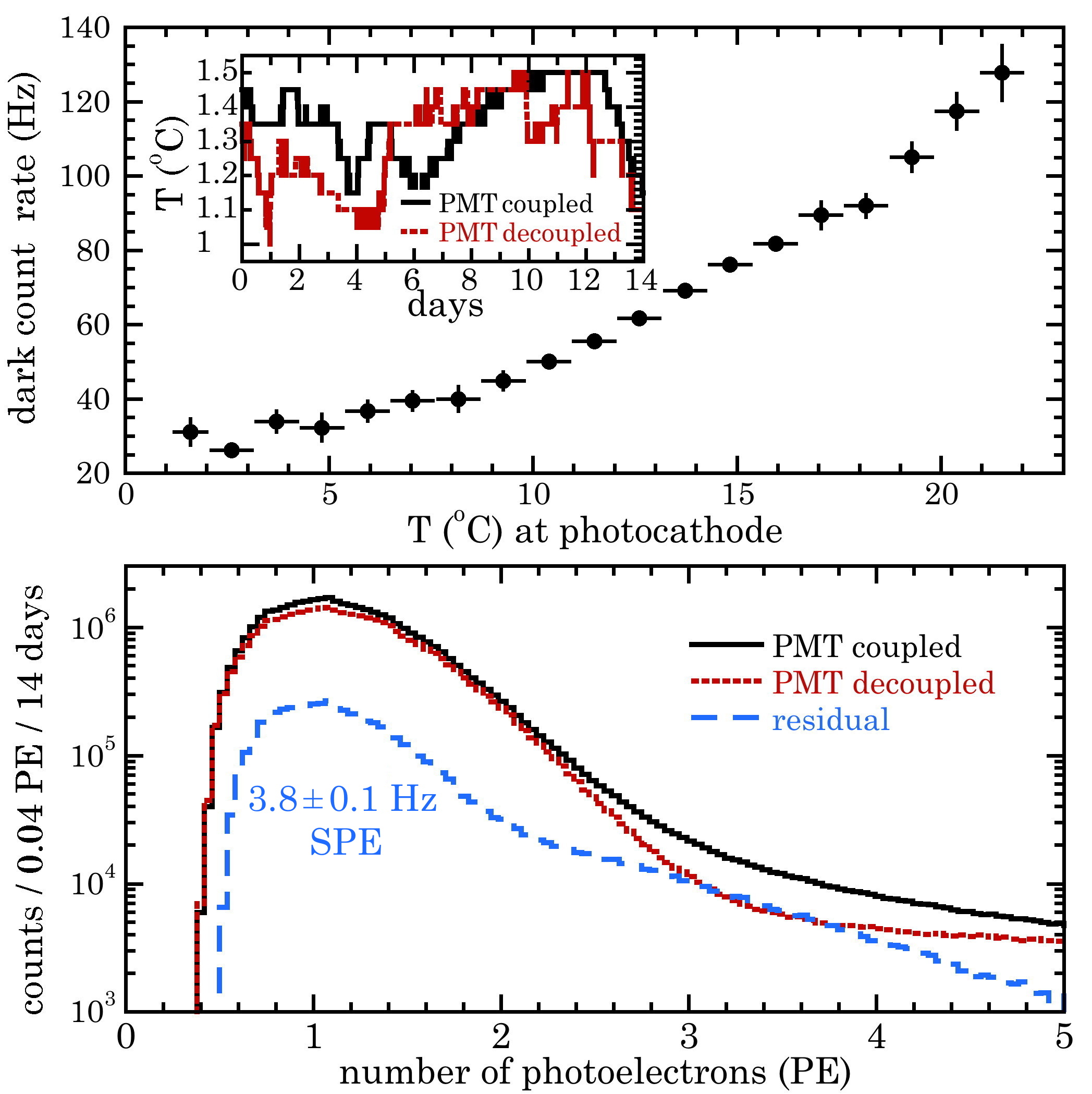}
\caption{\label{fig:lee}  {\it Top inset:}  stability of photocathode temperature achieved during preliminary single cooled detector runs (see text). An improved control of room temperature stability has been recently  achieved, further reducing the small fluctuations visible. {\it Top:} dependence of SPE emission rate on temperature, measured for ET9390QKA PMTs described in Sec.\ V. The dependence of this dark current on operating temperature  is minimized below few  $^{\circ}$C, a behavior characteristic of bialkali photocathodes \protect\cite{pmthdb1,pmthdb2}.  {\it Bottom:} residual SPE rate during preliminary cooled-PMT runs, previous to removal of ``slow" afterpulses (see text). A 0.11 Hz uncertainty arises from the temperature stability achieved (top panel).}
\end{figure}

The delayed-coincidence technique described in previous sections is able to probe unexplored SIMP parameter space for $m_{\chi}<$ 1 GeV, but is limited in its cross-section reach by the requirement that interactions take place in both cells,  the penalty imposed by the finite solid angle they mutually subtend, and by the rapid increase in background below a 3 PE threshold (Sec.\ V). An alternative approach relies on differential measurements using a single LS cell, cooled to $\sim1^{\circ}$C (Fig.\ 12) using an external circulating bath connected to an OFHC copper refrigeration line wrapped around the  cell flange. The DAQ system is modified to trigger on signals from this single cell, with a threshold at SPE level, as before. In this mode of operation, rates of SPE production with the PMT optically coupled to the LS cell, and decoupled by a thin foil of light-blocking material (i.e., with SPEs originating in PMT dark current only), are compared. Their difference is identified with the largest rate of SIMP interactions in the LS producing a SPE that is allowed by the data (Fig.\ 12). This approach is feasible as long as the temperature of the PMT photocathode is maintained as constant as possible between measurements (inset of Fig.\ 12), to guarantee nearly-identical contributions from PMT dark current to both datasets. Cooling of the PMT drastically reduces the impact of temperature stability on the uncertainty in this residual SPE rate (Fig.\ 12).\\ 

The $\sim$30 Hz single-cell trigger rate at $1^{\circ}$C (Fig.\ 12) is modest enough to permit vetoing of ``slow" afterpulses \cite{aftslow1,aftslow2,aftslow3,aftslow4}, while generating only a minimal impact on live-time. These highly-delayed SPE afterpulses are observed in the present system to follow very large energy depositions (cosmic ray-traversal), appearing with a multiplicity of a few per primary.  A long gate of order 500 $\mu$s following a trigger is required for their complete elimination. Minor modifications to the electronics setup of Fig.\ 5, necessary to remove this dominant background, will be described in  \cite{common}. Present experimentation indicates that a limit on the rate of SPE production by low-mass SIMPs of order 0.1 Hz is within reach, following  removal of slow afterpulses. Further progress would require a photocathode temperature stability much better than $0.1^{\circ}$C, and/or an additional reduction in PMT dark current. 

\begin{widetext}
\begin{figure*}[!htbp]
\includegraphics[width=1\textwidth]{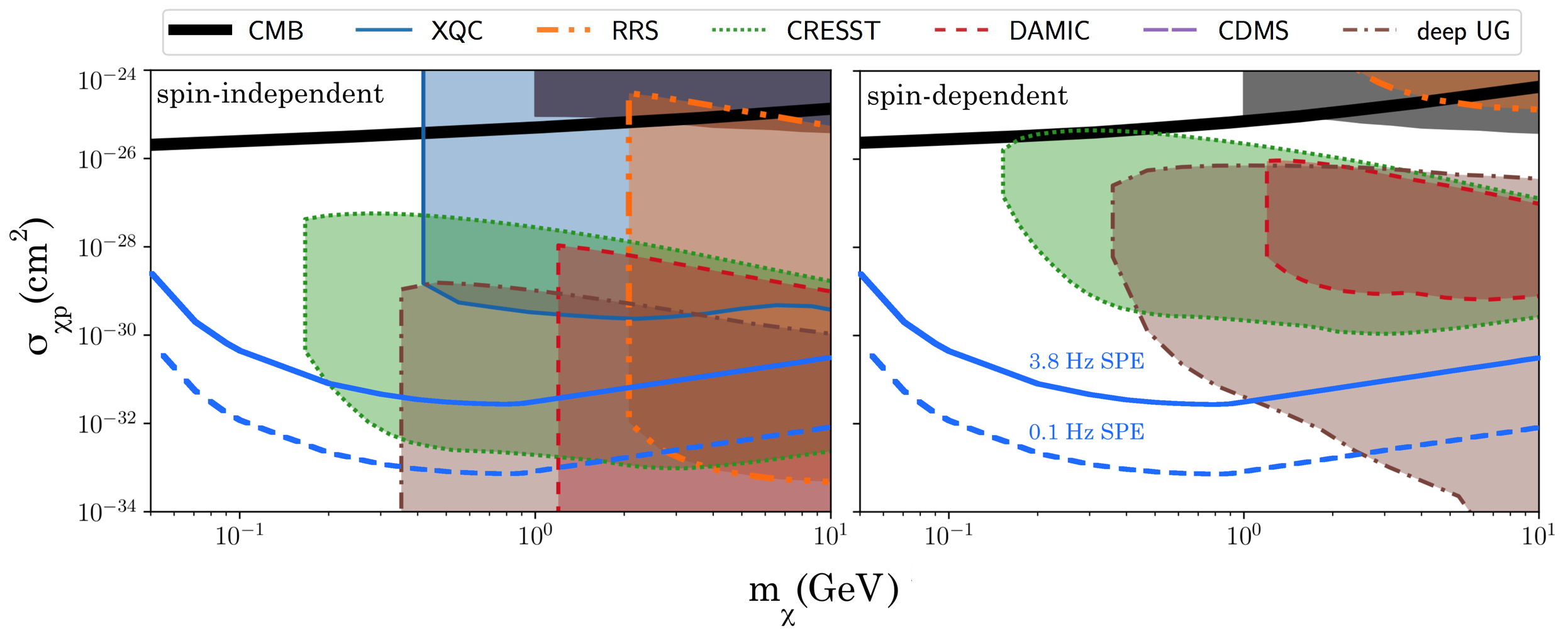}
\caption{\label{fig:lee} Solid blue line: present sensitivity to spin-dependent and -independent SIMP scattering off protons, derived from the preliminary runs in Fig.\ 12. Dashed blue line: expected improved sensitivity following a two-week exposure of the single cooled LS cell described in the text. Excluded regions from other searches are  from \protect\cite{hands}. The loss of sensitivity due to SIMP energy loss in the overburden (top boundary of closed contours) is not calculated for this search, but should be close to that from a recent  CRESST micro-bolometer run (green contour, \protect\cite{nucleus}), also performed in a surface lab. These contours assume preferential DM couplings to protons, with very similar regions existing when comparable couplings to neutrons are considered \protect\cite{hands}. A recently proposed process, nuclear-recoil bremsstrahlung \protect\cite{int14}, may further expand the reach of all techniques shown. The advantages of hydrogenated scintillators for low-mass SIMP searches, described in Sec.\ III, are made evident in this figure. }
\end{figure*}
\end{widetext}

For an interesting range of SIMP interaction cross-sections, the fast scintillation properties of EJ-301 would allow to look for indications of their multiple-scattering within a single large LS cell, separated by a characteristic TOF between scatters. For the small signals expected from low-mass SIMPs, this possibility is unfortunately encumbered  by  anomalous photoelectron trajectories within the PMT \cite{pmthdb1,smirnov,barros}, which are observed to generate occasional SPE afterpulsing within tens of ns from a primary SPE.

Fig.\ 13 translates the achievable 0.1 Hz SPE residual rate into  nuclear scattering cross-section vs.\ SIMP mass projected constraints. To arrive at the sensitivity shown in the figure, the rate of detectable signals from SIMP-proton scattering in a single LS cell is approximated as $R \approx \kappa \cdot \Phi(m_{\chi}) \cdot \sigma \cdot \varepsilon(m_{\chi},E_{th}) \cdot V$, where $\kappa=4.82\cdot10^{22}$ cm$^{-3}$ is the number density of hydrogen nuclei in EJ-301,  $\Phi\simeq\frac{\rho}{m_{\chi}}<\!\!v\!\!>$ is the SIMP flux through the  cell,  $\rho \simeq 0.3 ~$GeV/cm$^{3}$ is the commonly accepted local galactic DM mass density,  $<\!\!v\!\!>=335$ km/s is the average galactic DM speed at Earth,  $\sigma$ is the scattering cross-section, $\varepsilon(m_{\chi},E_{th})$ represents the efficiency in creating a signal above a scintillation threshold $E_{th}\geq1$ PE, and $V=1,500$ cm$^{3}$ is the active cell  volume.

The efficiency $\varepsilon$ is found via Monte Carlo simulation as follows: SIMP speed is sampled from its SHM distribution in the frame of reference of the Earth \cite{halo1,halo2,halo3,halo4}, defining its kinetic energy $T_{0}$. The proton recoil energy imparted is selected in an isotropic scattering approximation, i.e., assigning equal probability up to a maximum recoil energy of $4\frac{M\cdot m_{\chi}}{(M + m_{\chi})^{2}}T_{0}$, where $M=$ 0.938 GeV is the proton mass. This proton energy is translated into an expected mean number of PE via the modified Lindhard model in Fig.\ 3, and the measured EJ-301 light yield of 2.5 PE/keVee (a small favorable increase by $\sim$10\% at $1^{\circ}$C \cite{ejt} is neglected). The number of PE detected is sampled assuming Poisson fluctuations around this mean. The fraction of simulated events generating $\geq1$ PE is identified with  $\varepsilon$. As a reference, $\varepsilon=1.1\cdot10^{-3}$ (0.17) for $m_{\chi}= $ 0.1 (1) GeV. It should be noted that the adequacy of  Poisson statistics to describe the generation of scintillation by sub-keV protons in organic scintillator has  been  experimentally ascertained in \cite{ahlen1,ahlen2} (Fig.\ 3).

Fig.\ 13 clearly illustrates the advantages of hydrogenated organic scintillators for SIMP searches. However, their use is not limited to interactions mediated by nuclear recoils. The subject of DM interactions via electron scattering is of relatively new interest, with recent limits derived from XENON-10, SENSEI and SuperCDMS data only \cite{essig,sensei,scdms}. These extend down to a DM mass of few MeV, due to the absence of a quenching factor, and the possibility of a larger momentum transfer to electron targets. A similar expansion in sensitivity down to m$_{\chi}\sim$ 1 MeV is expected from the present search when considering this other interaction mechanism, as the minimum SIMP kinetic energy necessary for the production of a scintillation photon in organic scintillator is of just a few eV. The advantages of scintillators for sub-GeV dark matter detection have been recently emphasized \cite{derenzo}.

To finalize, additional applications of the delayed-coincidence method can be listed. For instance, a broad class of models predicts the possibility of DM capable of internal inelastic excitation. In these scenarios, highly-characteristic signals composed of a nuclear recoil in the first detector, followed by  de-excitation via low-energy gamma emission in the second, would be expected \cite{int12}. As already mentioned above, the delayed-coincidence method seems particularly well-adapted to explore such possibilities, constraining the parameter space (coupling, lifetime of the excitation) over which this mechanism might be responsible for the long-standing DAMA/LIBRA anomaly \cite{int12}. \\

\section{Conclusions}

WIMP searches are rapidly exhausting the range of possibilities left open for their original motivation, a lightest supersymmetric partner of cosmological relevance. Surprisingly, vast regions of SIMP parameter space have survived without dedicated exploration during this long period of concentration on WIMP searches, performed at depth. Experimentation with hydrogenated scintillators operated in a shallow underground site can probe some of these still viable dark matter candidates. The ability to achieve suitably-low background rates has been demonstrated in this work, following the identification and abatement of initially dominant sources.  Improved bounds on several possible mechanisms of SIMP interaction, over a broad range of SIMP masses, are expected from this ongoing effort \cite{common}. This reach is illustrated by the first experimental limits on dark matter candidates with a mass below 100 MeV, interacting preferentially via nuclear recoils. \\

\begin{acknowledgments}
~I am indebted to John Beacom, Christopher Cappiello, Joakim Edsj\"o, Glennys Farrar, Dan Hooper, Alexander Kavner, Chris Kouvaris, Sam McDermott, Ian Shoemaker, and Neal Weiner for useful exchanges. My gratitude also goes to Chuck Hurlbut and Chris Maxwell at Eljen Technology for their care in the development of low-background LS cells, and to Andy Cormack and Paul Davison at ET Enterprises for their patient selection of optimized PMTs. The Berglund Construction company is thanked for maintaining access and working conditions in the underground LASR laboratory during building upgrades. COHERENT work at the University of Chicago is supported through award NSF PHY-1506357. This work was also supported in part by the Kavli Institute for Cosmological Physics at the University of Chicago through grant NSF PHY-1125897, and an endowment from the Kavli Foundation and its founder Fred Kavli. This work is dedicated to the memory of Jim Cronin and Robert Metz.

\end{acknowledgments}

\end{document}